\documentclass[amsmath,amssymb,aps,prl,twocolumn,showpacs]{revtex4}
\usepackage{graphicx,psfrag,calc}

\begin{document}

\title{Density Induced Quantum Phase Transitions in Triplet Superconductors}

\author{R. W. Cherng}
\affiliation{Department of Physics, 
Harvard University,
Cambridge, MA 02138}

\author{C. A. R. S\'{a} de Melo}
\affiliation{School of Physics, Georgia Institute of Technology,
             Atlanta Georgia 30332}

\date{\today}

\begin{abstract}
We consider the possibility of quantum phase transitions in the
ground state of triplet superconductors where particle density is 
the tunning parameter.
For definiteness, we focus on the case of one band 
quasi-one-dimensional triplet superconductors but many of our conclusions 
regarding the nature of the transition are quite general. 
Within the functional integral formulation, we calculate the electronic compressibility 
and superfluid density tensor as a function of the particle density for 
various triplet order parameter symmetries
and find that these quantities are
non-analytic when a critical value of the particle density is reached.
\end{abstract}

\maketitle

Triplet superconductivity is a very rare, but very rich phenomenon in condensed matter physics.
The very few confirmed examples in nature include strontium ruthenate~\cite{maeno-94} (a Ruthenium oxide)
and the Bechgaard salt ${\rm (TMTSF)_2 PF_6}$~\cite{lee-02} (an organic molecular compound).
Because the full confirmation of triplet superconductivity in solids has occurred only over the
last few years, these lattice systems are not yet as fully studied theoretically and experimentally 
as $^{3}{\rm He}$, their neutral liquid superfluid counterpart~\cite{leggett-75}. 
Unlike $^{3}{\rm He}$, these systems are lattice charged superfluids, and their order parameters
are intimatelly related to the lattice periodicity as in d-wave 
high critical temperature ($T_c$) superconductors~\cite{dagotto-94}. 

\begin{figure}
\psfrag{N}[c][c]{$\widetilde{N}$}
\psfrag{V}[c][c]{$|V_0/t_x|$}
\psfrag{m1}[r][c]{\small $\mu=\mu^*_1$}
\psfrag{m2}[r][c]{\small $\mu=\mu^*_4$}
\begin{center}
\(
  \begin{array}{c}
   \multicolumn{1}{l}{\hspace{-0.2cm}\mbox{\bf (a)}} \\
     \includegraphics[width=7.0cm]{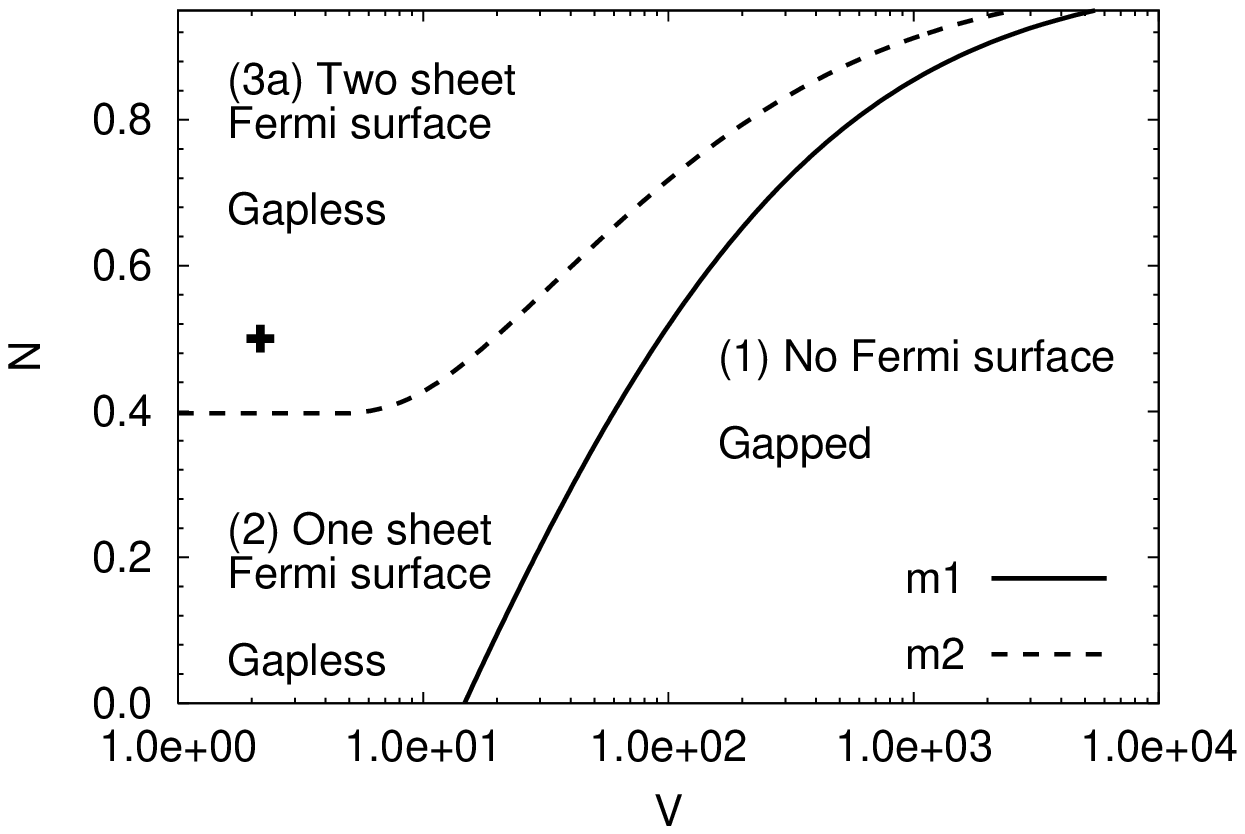} \\
    
   \multicolumn{1}{l}{\hspace{-0.2cm}\mbox{\bf (b)}} \\

\includegraphics[width=7.0cm]{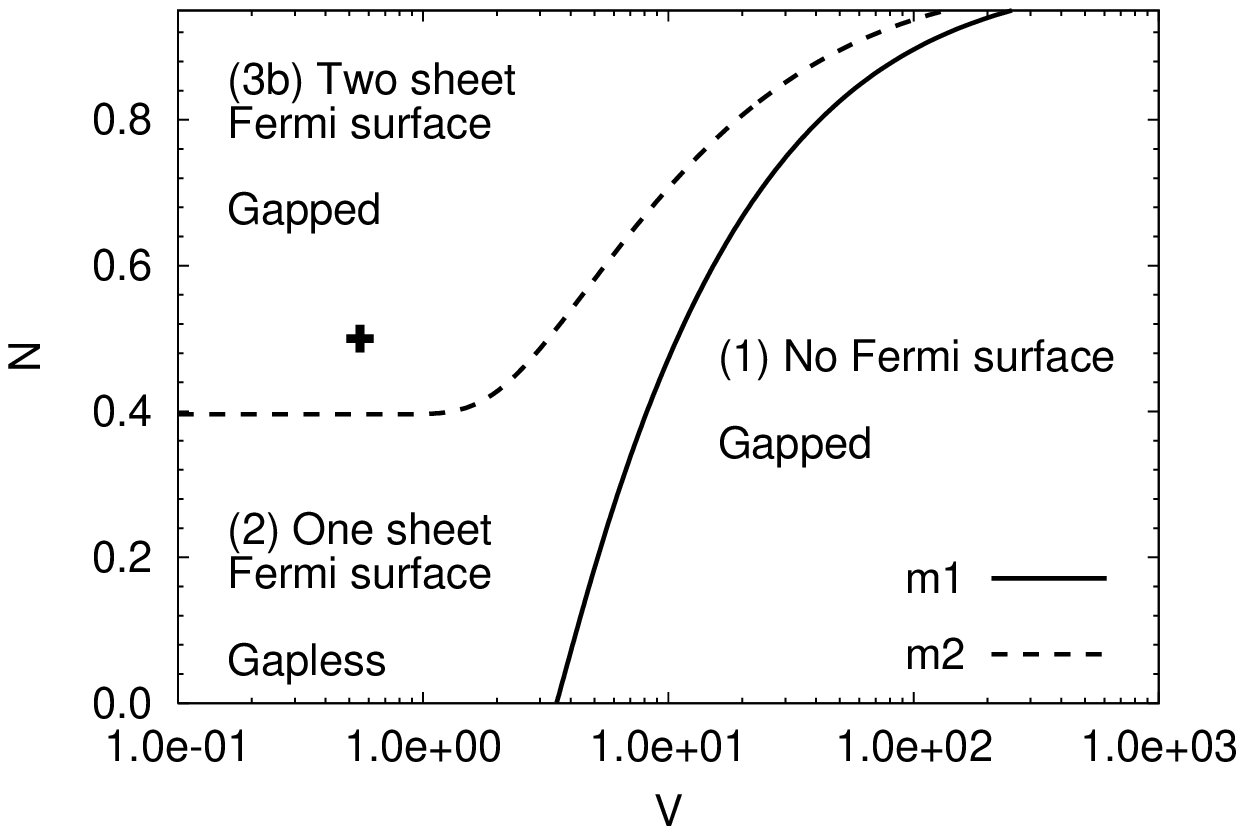}
\end{array}\)
\end{center}
\caption{
Phase diagrams for (a) $f_{xyz}$ (b) $p_x$ symmetries based on 
Fermi surface connectivity
and excitation spectrum.  
The phase diagrams are symmetric (not shown) around half-filling (${\widetilde N} = 1$) due
to particle-hole symmetry. The small cross indicates the parameters compatible to the known
triplet superconductor ${\rm (TMTSF)_2 PF_6}$ 
${\widetilde N} = 0.5$ and $V_{p_x}/t_x = 0.5531$ for $p_x$, ${\widetilde N} = 0.5$ and
$V_{f_{xyz}}/t_x = 2.1685$. }
\label{fig:one}
\end{figure}

It is known experimentally that electronic properties and ground states of cuprate superconductors
(d-wave singlet)~\cite{dagotto-94} and Strontium Ruthenate (p-wave triplet)~\cite{mackenzie-03}
are very sensitive to chemical doping
while these properties for ${\rm (TMTSF)_2 PF_6}$ (p-wave triplet)~\cite{lee-02} are very 
sensitive to both external pressure and chemical doping. However, recent experiments have
demonstrated that it is possible to change the carrier density electrostatically 
in cuprate superconductors~\cite{ahn-03} and amorphous Bismuth~\cite{goldman-05}
without the introduction of additional disorder that often occurs for chemical doping.
It is likely that similar electrostatic 
techniques will be developed for use in other superconductors like Strontium Ruthenate or 
${\rm (TMTSF)_2 X}$ (with ${\rm X = ClO_4, PF_6}$).  Since it may be possible in the near 
future to tune the carrier concentration with the use of field effect techniques, some important
theoretical questions concerning these three types of systems 
may soon receive an experimental answer. For instance, 
are there quantum critical points separating magnetic and 
superconducting order as a function of particle density?
Or, are there quantum critical points within the superconducting phase
as a function of particle density?
In Fig.~\ref{fig:one} we show the zero temperature density versus interaction 
phase diagram indicating the existence of quantum critical lines, where the order paramater
does not change symmetry but the ground state topology changes. The elementary excitation spectrum
also changes from gapless to gapped, and at finite temperatures there are three distinct
regions the Fermi liquid, the pseudo-gap, and the Bose liquid regions.

In anticipation of experimental efforts, we propose to study the possible existence 
of topological quantum phase transitions in lattice triplet superconductors as
a function of particle density.
We focus on the specific case of the ${\rm (TMTSF)_2 X}$ family, because they are 
a single-band triplet superconductors, unlike Strontium Ruthenate where three bands may be necessary 
to describe triplet superconductivity~\cite{mackenzie-03}. 
For this purpose, we study single band quasi-one-dimensional 
systems in an orthorhombic lattice with dispersion 
\begin{equation}
\label{eqn:dispersion}
\epsilon_{\bf k} = -t_x \cos(k_x a) -t_y \cos(k_y b) -t_z \cos(k_z c),
\end{equation}
where ${t_x} \gg {t_y} \gg {t_z}$.
We work with the Hamiltonian
$
H = H_{kin} + H_{int},
$
where the kinetic energy part is 
$
H_{kin} = \sum_{{\bf k},\alpha} \xi_{\bf k}
\psi_{{\bf k}, \alpha}^{\dagger} \psi_{{\bf k}, \alpha},
$
with $\xi_{\bf k} = \epsilon_{\bf k} - \mu$, 
where $\mu$ may include the Hartree shift. 
The interaction part is
\begin{equation}
\label{eqn:hint}
H_{int} = 
\dfrac{1}{2} \sum_{{\bf k} {\bf k^{\prime}} {\bf q}} 
\sum_{\alpha \beta \gamma \delta}
V_{\alpha \beta \gamma \delta} ({\bf k}, {\bf k^{\prime}})
b_{\alpha \beta}^{\dagger} ({\bf k}, {\bf q})
b_{\gamma \delta} ({\bf k^{\prime}}, {\bf q})
\end{equation}
with
$
b_{\alpha \beta}^{\dagger} ({\bf k}, {\bf q}) = 
\psi_{-{\bf k} + {\bf q}/2, \alpha}^{\dagger}
\psi_{{\bf k} + {\bf q}/2, \beta}^{\dagger},
$
where the labels $\alpha$, $\beta$, $\gamma$ and $\delta$ 
are spin indices and 
the labels ${\bf k}$, ${\bf k}^{\prime}$ and ${\bf q}$ represent 
linear momenta. We use units where $\hbar = k_B = 1$. 
In the case of weak spin-orbit coupling
and triplet pairing, the model interaction tensor can be chosen to be
\begin{equation}
V_{\alpha \beta \gamma \delta} ({\bf k}, {\bf k^{\prime}})
= - \vert V_{\Gamma} \vert 
h_{\Gamma} ({\bf k}, {\bf k^{\prime}})
\phi_{\Gamma} ({\bf k}) \phi^{*}_{\Gamma} ({\bf k^{\prime}})
\Gamma_{\alpha \beta \gamma \delta},
\end{equation}
where 
$\Gamma_{\alpha \beta \gamma \delta} = {\bf v}_{\alpha \beta} \cdot
{\bf v}_{\gamma \delta}^{\dagger}/2$ with 
${\rm v}_{\alpha \beta} = (i\sigma \sigma_y)_{\alpha \beta}$. 
$V_{\Gamma}$ is a prefactor with dimensions of energy which characterizes 
a given symmetry. Furthermore, the term 
$h_{\Gamma} ({\bf k}, {\bf k^{\prime}}) 
\phi_{\Gamma} ({\bf k}) \phi^{*}_{\Gamma} ({\bf k^{\prime}})$ 
contains the momentum and symmetry dependence of the interaction 
of the irreducible representation $\Gamma$ with basis function
$\phi_{\Gamma} ({\bf k})$ and 
$\phi_{\Gamma}^{*} ({\bf k^\prime})$ 
representative of the orthorhombic group ($D_{2h}$).  

We write down the  partition function~\cite{sdm-97} 
$Z = \int {\cal D} \left[\psi^\dagger, \psi \right] {\rm exp} 
\left[ - S \right] $
with the action $S = \int_0^\beta d\tau
\left[
\sum_{{\bf k}, \alpha} \psi_{\bf k}^{\dagger} (\tau) (\partial_\tau ) 
\psi_{\bf k} (\tau) + H (\psi^{\dagger}, \psi) 
\right].$
For simplicity, we take 
$h_{\Gamma} ({\bf k}, {\bf k^{\prime}}) = 1 $.
We introduce the vector order parameter for triplet superconductivity
through the Hubbard-Stratanovich transformation and integrate out
the fermions to obtain the effective action
\begin{equation}
\label{eqn:eff-action}
S_{\rm eff} = 
Q_{G} + \int_0^\beta d\tau \sum_{{\bf k} \alpha} \frac{\xi_{k}}{2}
- \beta^{-1} {\rm Tr} \ln \left[ \frac{\beta{\bf M}}{2} \right],
\end{equation}
where 
$
Q_G = \int_0^\beta d\tau \sum_{{\bf q}, i} {\cal D}_i^\dagger ({\bf q}, \tau)
{\cal D}_i ({\bf q}, \tau) /\vert V_\Gamma \vert
$
and
${\bf M}$ is the matrix
$$
{\bf M} = 
\left(
\begin{array}{cc}
\left[ \partial_{\tau} + \xi_{\bf k} \right] \delta_{\bf k_1 k_2} \delta_{\alpha \beta} & 

{\bf A} ({\bf k_1}, {\bf k_2}, \tau) \\

{\bf A}^\dagger ({\bf k_1}, {\bf k_2}, \tau) &

\left[ \partial_{\tau} - \xi_{\bf k} \right] \delta_{\bf k_1 k_2} \delta_{\alpha \beta} \\ 
\end{array}
\right),
$$
where ${\bf A} = 
 -\sum_i \phi_{\Gamma} ( { {\bf k_1} - {\bf k_2} \over 2} ) {\cal D}_{i} ( {\bf k_1} + {\bf k_2}, \tau ) 
{\rm \bf  v}_{\alpha \beta, i}$.
At the saddle point,
${\cal D}_{i} ({\bf k_1} + {\bf k_2}, \tau)$ is taken to be 
$\tau$ independent, and to have total Fermion center of mass momentum ${\bf k_1} + {\bf k_2} = 0$.
Thus, 
$
{\cal D}_{i} ({\bf k_1} + {\bf k_2}, \tau) = 
\eta_i \Delta_{\Gamma} \delta_{ {\bf k_1} + {\bf k_2}, 0} + 
\delta {\cal D}_{i} ({\bf k_1} + {\bf k_2}, \tau). 
$

The order parameter equation is obtained from the 
stationary condition $\delta S_{\rm eff}^{(0)}/\delta {\cal D}_{i}^{\dagger} 
= 0$, where $S_{\rm eff}^{(0)}$ is the saddle point action 
\begin{equation}
\label{eqn:order-parameter}
1 = \sum_{\bf k} \vert V_{\Gamma} \vert |\phi_{\Gamma} ({\bf k})|^2 
\tanh (\beta E_{\bf k}/2) / 2 E_{\bf k},
\end{equation}
where 
$
E_{\bf k} = 
\sqrt{ \xi_{\bf k}^2 + |\Delta_{\Gamma}|^2 |\phi_{\Gamma} ({\bf k})|^2},
$
corresponds to the quasiparticle excitation energy.
The number equation is obtained from 
$N = - \partial \Omega/\partial \mu$, 
where $\beta \Omega  = - \ln Z$ is
the thermodynamic potential,
\begin{equation}
\label{eqn:number}
N = N_0 + N_{\rm fluct},
\end{equation}
where $N_0 = 
\sum_{\bf k} n_{\bf k}$, 
and $n_{\bf k} = 
\left[
1 - \xi_{\bf k} \tanh (\beta E_{\bf k}/2) / E_{\bf k} 
\right]
$
is the momentum distribution.
The additional term $N_{\rm fluct} = - \partial \Omega_{\rm fluct}/\partial \mu$, 
where $\Omega_{\rm fluct}$ are Gaussian fluctuations to saddle point $\Omega_0$.
These two equations must be solved self-consistently, and 
quite generally they are correct even in the strong coupling (or low density)
regime provided that $T \ll T_c$~\cite{sdm-97}, where
$N_{\rm fluct} \sim T^4$ for all couplings~\cite{footnote}.

The saddle point vector $ {\cal D}_i^{(0)} = \eta_i \Delta_\Gamma$ is related to the standard ${\bf d}$-vector
via the relation 
$
{d_i} ({\bf k}) = 
\sum_{\bf k^\prime} {\cal D}_i^{(0)} \phi_{\Gamma} ({\bf k}).
$
In the $D_{2h}$ point group all representations are one dimensional 
and non-degenerate~\cite{duncan-01}, which means that
the ${\bf d}$-vector in momentum space for unitary triplet
states in the weak spin-orbit coupling limit
is characterized by one of the four states:
(1) $^3 A_{1u} (a)$, 
with  
${\bf d} ({\bf k}) = {\hat \eta} \Delta_{f_{xyz}} X Y Z$ 
(``$f_{xyz}$'' state);
(2) $^3 B_{1u} (a)$, 
with  
${\bf d} ({\bf k}) = {\hat \eta} \Delta_{p_z} Z$ (``$p_{z}$'' state);
(3) $^3 B_{2u} (a)$, 
with  
${\bf d} ({\bf k}) = {\hat \eta} \Delta_{p_y} Y$ (``$p_{y}$'' state);
(4) $^3 B_{3u} (a)$, 
with  
${\bf d} ({\bf k}) = {\hat \eta} \Delta_{p_x} X$ (``$p_{x}$'' state).
Since, the Fermi surface can touch
the Brillouin zone boundaries the functions 
$X$, $Y$, and $Z$ need to be periodic and can 
be chosen to be $X = \sin{(k_{x} a)}$, 
$Y = \sin{(k_{y} b)}$, 
and $Z = \sin{(k_{z} c)}$.
The unit vector $\hat \eta$ defines the direction 
of ${\bf d} ({\bf k})$.
From here on we scale all energies by $t_x$.
The parameters used are 
$t_x = 1.0000$, 
$t_y = 0.2144$ and 
$t_z = 0.0083$. 
It will be easier to discuss the properties of the Fermi surface and
quasiparticle excitation spectrum by considering the chemical potential $\mu$ 
instead of the filling factor ${\widetilde N}$.  
Consider the ``normal state'' Fermi surface defined in the first Brillouin zone (BZ) by  
$\xi_k=0$, keeping in mind the periodicity in $k$-space.  Let us define 
the following characteristic values $\mu^*_1\equiv t_x+t_y+t_z$,
$\mu^*_2\equiv t_x+t_y-t_z$,
$\mu^*_3\equiv t_x-t_y+t_z$,
$\mu^*_4\equiv t_x-t_y-t_z$.
For $\mu<\mu^*_1$ the chemical potential is below the bottom
of the band and there is no Fermi surface.  
For $\mu^*_1<\mu<\mu^*_2$ the Fermi surface consists of one connected sheet contained entirely 
in the first BZ and is topologically equivalent to a sphere of genus zero.
For $\mu^*_2<\mu<\mu^*_3$, the Fermi surface is one connected sheet touching the edges
of the BZ on the planes  $k_z=\pm\pi$ equivalent to a torus of genus one.
For $\mu^*_3<\mu<\mu^*_4$, the Fermi surface is one connected sheet touching the edges
of the BZ on the planes $k_z=\pm\pi$ and $k_y=\pm\pi$ equivalent to a torus of genus two.
Lastly for $\mu^*_4<\mu$, the Fermi surface consists of two disconnected sheets each of which
touch the BZ boundary and is equivalent to two disconnected tori each of genus one. 

For the superconducting state, the intersection of the Fermi surface and order parameter nodes constitute
the loci of gapless quasiparticle excitations.  For the $p_i$ symmetry (where $i$ is $x$, $y$, or $z$) 
the order parameter nodes are on the planes $k_i=0,\pm\pi$.
For the $f_{xyz}$ symmetry, the order parameter nodes are on the union of the planes
$k_i=0,\pm\pi$, where $i$ is $x$, $y$, and $z$.
For all symmetries, the quasiparticle excitations are fully gapped for $\mu< \mu^*_1$ since there is no Fermi
surface. For $\mu^*_1<\mu<\mu^*_4$, the order parameter nodes for all symmetries intersect the 
Fermi surface and hence quasiparticle excitations are gapless.
For $\mu^*_4<\mu$ the Fermi surface splits into two sheets that separate along $k_x$ so that
the $p_x$ nodes no longer intersect the Fermi surface opening a gap in the quasiparticle excitation spectrum. 
However, the $f_{xyz}$, $p_z$, and $p_y$ nodes still intersect the Fermi surface so quasiparticle excitations
remain gapless.

We plot representative phase diagrams for the $f_{xyz}$  and $p_x$ symmetries in Fig.
\ref{fig:one}.  From the discussion above, the $p_y$ and $p_z$ phase diagrams
are qualitatively similar to the $f_{xyz}$ phase diagrams.  There are three distinct phases 
characterized by Fermi surface connectivity and quasiparticle excitation spectrum:
(1) no Fermi surface and fully gapped $E ({\bf k})$ for all symmetries ($\mu<\mu^*_1$),
(2) one sheet Fermi surface and gapless for all symmetries ($\mu^*_1<\mu<\mu^*_4$),
(3a) two sheet Fermi surface and gapless for $f_{xyz}$, $p_z$, $p_y$  ($\mu^*_4<\mu$)
(3b) two sheet Fermi surface and fully gapped for $p_x$ ($\mu^*_4<\mu$).
In addition, the (2) phase splits into three regions using the finer classification of 
Fermi surface topological genus:
(2i) genus zero ($\mu^*_1 < \mu < \mu^*_2$), (2ii) genus one ($\mu^*_2 < \mu < \mu^*_3$),
(2iii) genus two ($\mu^*_3 < \mu <\mu^*_4$).
There are several qualitative features of interest in the phase diagrams.  For a fixed density
and all symmetries, the chemical potential does not go below the bottom of the band ($\mu < \mu_1^*$)
until a critical coupling is reached.
This critical coupling is
$V_{f_{xyz}}/t_x = 14.8021$,
$V_{p_z}/t_x = 1.8150$,
$V_{p_y}/t_x = 2.3952$,
$V_{p_x}/t_x = 3.5052$ for ${\widetilde N} = 0$.
In addition, consider the approach to the strict one-dimensional 
limit ($t_y,t_z\rightarrow 0$).  If $t_z\rightarrow 0$, the $\mu^*_2$ ($\mu^*_3$) boundary
merges with $\mu^*_1$ ($\mu^*_4$) leaving only the (2ii) region.  
If in addition $t_y\rightarrow 0$, the $\mu^*_1$ boundary merges with $\mu^*_2$ producing only
one boundary between the (1) and (3) phases.

\begin{figure}
\begin{center}
\psfrag{N}[c][c]{$\widetilde{N}$}
\psfrag{k}[c][c]{$\kappa|t_x|$}
\includegraphics[width=7.0cm]{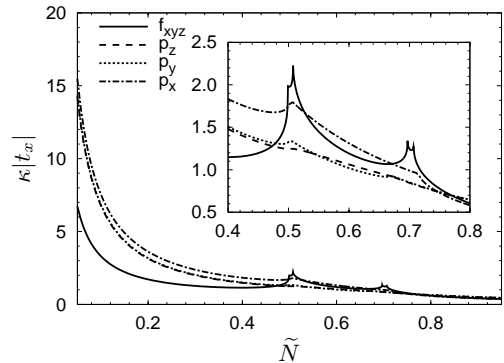} \\
\end{center}
\caption{
Filling factor (${\widetilde N}$) dependence of the dimensionless electronic compressibility $\kappa|t_x|$.  The
inset shows the region ${\widetilde N} = 0.4-0.8$.}
\label{fig:two}
\end{figure}
We now turn our attention to thermodynamic quantities that provide signatures of the
topological changes discussed above.  In the following calculations we fix the interaction strength to be
$V_{f_{xyz}}/t_x = 91.7202$,
$V_{p_z}/t_x = 11.8747$,
$V_{p_y}/t_x = 11.7781$,
$V_{p_x}/t_x = 10.8297$,
which forces $\mu=\mu^*_1$ at ${\widetilde N}=0.5$.  In contrast, $\mu=\mu^*_4$ at
${\widetilde N} =0.706$, ${\widetilde N} =0.674$, ${\widetilde N} = 0.672$, ${\widetilde N} =0.719$ 
for the $f_{xyz}$, $p_z$, $p_y$, $p_x$ symmetries, respectively.  
The $T = 0$ electronic compressibility $\kappa=N^{-2} (\partial N/ \partial \mu)_{T, V}$ is
\begin{equation}
\label{eq:kappa}
\kappa=\frac{2}{N^2}\sum_k\left[\frac{1}{2E_k}\left(1-\frac{\xi_k^2}{E_k^2}\right)\right].
\end{equation}
which we plot in Fig. \ref{fig:two}.
Although the compressbility does not formally diverge, there are clear anomalies (non-analyticities) when
the Fermi surface topology and quasiparticle excitation spectrum change.  As ${\widetilde N}$ increases, $\mu$ increases
and crosses the boundaries $\mu^*_j$ where the first derivative of $\kappa$ decreases
discontinuously.  Within the $p$-symmetries, the magnitude of this jump is largest for $p_x$.  
The $f_{xyz}$ symmetry, with the presence of
double or even triple nodes compared to the single nodes for the $p$ symmetries, has 
the largest jumps in the derivative of $\kappa$ clearly identified as the four cusps in Fig. 
\ref{fig:two}. These non-analyticities in $\kappa$ at $T = 0$ are indicative of a quantum phase
transition. At finite temperatures the cusps in $\kappa$ are smeared-out, but clear peaks are still present 
so long as one remains in the quantum critical region. The measurement of the electronic compressibility
may be achieved in a field effect geometry~\cite{ahn-03, goldman-05} through the relation 
\begin{equation}
\label{eqn:kappa-capacitance}
\kappa = V C_d/Q^2,
\end{equation}
where $C_d = [\partial Q / \partial V_e ]_{T, V}$ is the differential capacitance, 
$V_e$ is the applied voltage, $Q$ is the absolute value
of the total charge of carriers, and $V$ is the sample volume. 

\begin{figure}[floatfix]
\psfrag{N}[c][c]{$\widetilde{N}$}
\psfrag{rzz}[c][c]{$V_0\rho_{zz}/a_z^2|t_x|$}
\begin{center}
\includegraphics[width=7.0cm]{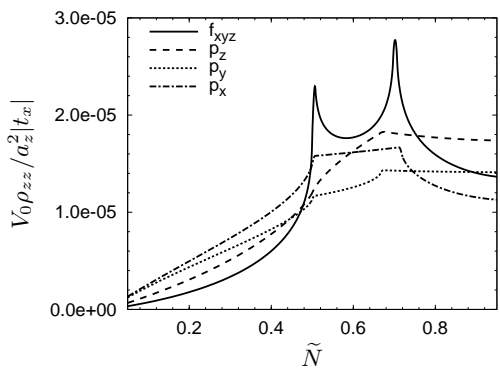} \\
\end{center}
\caption{
Filling factor (${\widetilde N}$) dependence of the $zz$ component of the dimensionless 
superfluid density tensor $V_0 \rho_{zz}/a_z^2|t_x|$ for various symmetries. 
$V_0$ $(a_z)$ is the unit cell volume (length).}
\label{fig:three}
\end{figure}

Next, we analyse the effect of phase fluctuations given by the action
\begin{equation}
\label{eqn:phase-action}
\Delta S  =  \frac {1}{8} \sum_{ {\bf q}, \omega_n } 
\left[ [ A(\omega_n)^2  + \rho_{ij} q_i q_j \right] \phi (q) \phi (-q), 
\end{equation}
where $A = N^2 \kappa/V$ is proportional to $\kappa$ and 
\begin{equation}
\label{eqn:superfluid-density}
\rho_{ij}= {1 \over V} 
\sum_{\bf k}
\left[ n_{\bf k}
{\partial_i} {\partial_j} \xi_{\bf k}
\right],
\end{equation}
is the superfluid density.
Here, $n_{\bf k}$
is the momentum distribution.
In Fig.~\ref{fig:three} we show
the ${\widetilde N}$ dependence of the $zz$ component of $\rho_{ij}$. 
In the case of the $D_{2h}$ group only diagonal components
$\rho_{ii}$ exist, but they are highly 
anisotropic due to the quasi-one-dimensionality
of $\xi_{\bf k}$.

From Eq. \ref{eqn:superfluid-density} it is clear that the zero-temperature
superfluid density is the curvature of the dispersion weighted by the momentum 
distribution.  We find the $xx$ component is a monotonically increasing function of
${\widetilde N}$ which is best understood as a consequence of quasi-one-dimensionality.  Neglecting the
curvature of the Fermi surface due to finite $t_x$, $t_y$, as ${\widetilde N}$ increases, the
Fermi surface encloses a monotonically increasing pocket of $k$-space around $k_x=0$.
In addition, below half filling the curvature of $\xi_k$ is of the same sign in this regime
so there is little cancellation between different regions of $k$-space.  
We find $\rho_{yy}$ is also smooth but peaks in
the region bounded by ${\widetilde N}\approx 0.5-0.8$ corresponding to $\mu\approx\mu^*_1-\mu^*_4$.
No longer neglecting the curvature of the Fermi surface due to finite $t_y$, we see
that for ${\widetilde N} \approx 0.5-0.8$, the Fermi surface encloses an increasing pocket of $k$-space
around $k_y = 0$ up to the edge of the BZ.
Since the curvature of $\xi_k$ changes sign at $k_y=\pi/2$, cancellations between 
different regions of $k$-space eventually occur.
The $zz$ component is the most interesting as it exhibits clear anomalies such as those seen
in $\kappa$.  The Fermi surface varies rapidly along the $k_z$ direction as a function of ${\widetilde N}$ so
that no simple analysis in terms of contributing regions of $k$-space is possible.  It appears
that the $\rho_{zz}$ is a more direct probe of the anomalies seen in $\kappa$ with clear
kinks as a function of ${\widetilde N}$. These non-analyticities 
again indicate the existence of a quantum phase transition as $\mu$ is tuned below the bottom of the band.

Lastly, it is easy to extract from $\Delta S$ the phase-only collective mode
frequencies via the substitution $i \omega_n  \to \omega + i \delta$. In this case $\omega ({\bf q}) = 
\sqrt { ( c_x^2  q_x^2 + c_y^2  q_y^2 + c_z^2  q_z^2 )}$,
where $c_x^2 = \rho_{xx}/A$, $c_y^2 =  \rho_{yy}/A$, and
$c_z^2 =  \rho_{zz}/A$. These collective mode frequencies, also show similar
anomalies to those of $\kappa$ and $\rho_{ij}$, and can be used to characterize the
quantum phase transition as well~\cite{footnote}. Notice that
$\omega ({\bf q})$ is anisotropic reflecting the orthorhombic structure.
For instance, $\omega (q_x, 0, 0) = c_x q_x$, 
$\omega (0, q_y, 0) = c_y q_y$, and  $\omega (0, 0, q_z) = c_z q_z$, where $c_i$ is the speed
of sound along the $i$-th direction.  However, these modes may be plasmonized in 
a charged superfluid.

In summary, we studied possible quantum phase transitions in triplet superconductors, 
as the density of carriers is changed, provided that the Cooper pairing interaction is
sufficiently attractive. 
For organic quasi-one-dimensional conductors (Bechgaard salts) only one quantum phase
transition may be accessible experimentally as the interaction strength is too weak
to cross both phase boundaries as a functions of filling factor (See Fig.~\ref{fig:one}).
However, these quantum phase transitions may be possible in optical lattices where
the attractive interaction may be changed via Feshbach resonances, thus allowing
the system to cross the phase boundary separating the weak and strong interaction regimes~\cite{esslinger-05}.
In order to identify these quantum phase transitions (QPT) where the symmetry of the order parameter
does not change, we classified Fermi surface topologies and excitation spectrum
properties as a function of filling factor for weak spin-orbit coupling symmetries 
$f_{xyz}$, $p_x$, $p_y$ and $p_z$.
We then related non-analyticities in the electronic compressibility and superfluid 
density to quantum phase transitions between various phases.
We would like to thank NSF (Grant No. 
DMR-0304380) and NDSEG for support.


\begin{thebibliography}{10}
\bibitem{footnote}
For low densities and strong interactions, Landau damping
is absent at zero temperature for all symmetries since
the elementary excitation spectrum is gapped.
At low densities and weak interactions,
the gap closes and there is Landau damping at low energies because
Cooper pairs near the nodal regions can decay into
the two-particle continuum.  However, the system
is still underdamped and the low temperature
behavior of $N_{\rm fluct}$ is controlled by $T^4$.
At higher densities and weak interactions,
the gap reopens in the $p_x$ symmetry and Landau
damping is absent while $p_y$, $p_z$, and $f_{xyz}$
symmetries remain gapless but underdamped.
\end{thebibliography}
\end{document}